\documentclass[aps,prl, preprintnumbers,amsmath,amssymb,latexsym,array,enumerate,letter,twocolumn,superscriptaddress]{revtex4-2}

\usepackage{amssymb}
\usepackage{amsmath}
\usepackage{epsfig}
\usepackage{hyperref}
\usepackage{breakurl}
\usepackage{xcolor}
\usepackage{slashed}

\usepackage{graphicx, subfigure, placeins, float}

\usepackage{extarrows}

\usepackage{enumitem,kantlipsum}

\graphicspath{{./figure/}}

\newcommand{\itbf}[1]{\textbf{\textit{#1}}}
\newcommand{\QQ}{{\bf q}}

\begin{document}

\preprint{
USTC-ICTS/PCFT-21-45}
\title{Double Copy of Form Factors and Higgs Amplitudes: \\
{\normalsize A Mechanism of Turning Spurious Poles in YM into Physical Poles in Gravity}}
\author{Guanda Lin}
\email{linguandak@pku.edu.cn}
\affiliation{CAS Key Laboratory of Theoretical Physics, Institute of Theoretical Physics, Chinese Academy of Sciences,  Beijing, 100190, China}
\author{Gang Yang}
\email{yangg@itp.ac.cn}
\affiliation{CAS Key Laboratory of Theoretical Physics, Institute of Theoretical Physics, Chinese Academy of Sciences,  Beijing, 100190, China}
\affiliation{School of Fundamental Physics and Mathematical Sciences, Hangzhou Institute for Advanced Study, UCAS, Hangzhou 310024, China}
\affiliation{International Centre for Theoretical Physics Asia-Pacific, Beijing/Hangzhou, China}
\affiliation{Peng Huanwu Center for Fundamental Theory, Hefei, Anhui 230026, China}

\begin{abstract}
We extend the double copy picture of scattering amplitudes to a class of matrix elements (so-called form factors) that involve local gauge invariant operators. Both the Bern, Carrasco and Johansson (BCJ) and the Kawai, Lewellen and Tye (KLT) formalisms are considered and novel properties are observed. One remarkable feature is that through the double-copy construction, certain spurious poles hidden in the gauge form factors become physical propagators in gravity. This mechanism also reveals new hidden relations for form factors which can be understood as a generalization of the BCJ relations. 

\end{abstract}

\maketitle

\section{Introduction}

\noindent
Despite the very different nature, gauge and gravity theories are known to be intimately related. 
In paricular, the perturbative amplitudes in gauge and gravity theories are related via the double copy as ``gravity = (gauge theory)$^2$'', realized in various formalisms including the
Kawai, Lewellen and Tye (KLT) relations \cite{Kawai:1985xq}, 
the Cachazo, He and Yuan (CHY) formula \cite{Cachazo:2013hca, Cachazo:2014xea} and especially the Bern, Carrasco and Johansson (BCJ) double copy stemming from the color-kinematics (CK) duality \cite{Bern:2008qj, Bern:2010ue} (see \cite{Bern:2019prr} for an excellent review).

Apart from scattering amplitudes that only involve  on-shell asymptotic states,
gauge invariant local operators also play important roles in gauge theories, and it is natural to ask:
does a consistent double-copy picture exist for physical quantities involving local operators?
Nevertheless, the answer is not obvious at all, since for example, local operators in gravity would break the diffeomorphism invariance.

In this paper, we make a concrete step towards addressing this question, 
by realizing both BCJ and KLT double copy for form factors.
Form factors (FFs) are 
defined as matrix elements 
between a gauge invariant operator $\mathcal{O}$ and $n$ on-shell states \cite{Maldacena:2010kp,Brandhuber:2010ad, Bork:2010wf}
(see \cite{Yang:2019vag} for a recent introduction and review),
\begin{equation}
    \mathcal{F}_{\mathcal{O},n} = \int d^{D} x e^{-i q \cdot x}\langle 1\, 2  \ldots  n |\mathcal{O}(x)| 0\rangle \,,
\end{equation}
where $q\ {=}\sum_{i=1}^n p_i$ is the off-shell momentum associated with the operator.
Although CK duality has been applied to compute high-loop FFs in gauge theories \cite{Boels:2012ew,Yang:2016ear, Lin:2020dyj, Lin:2021kht,Lin:2021qol}, it remains open problems about how to ``double-copy" those results and what the interpretations are in gravity theories. In our new double-copy realization, a careful consideration on the operator-induced relations is essential, which reveals several novel features.

One intriguing feature is that special \emph{spurious poles} appear in the construction of CK-dual numerators in gauge-theory FFs, and after double copy they become new \emph{physical} propagators in the gravity quantities, \emph{i.e.}
\begin{equation}
\textrm{spurious poles} \ 
\xlongrightarrow[\mbox{}]{\scriptstyle \textrm{double-copy}} \ 
\textrm{physical propagators} .
\nonumber
\end{equation}
Besides, the factorizations on the new propagators in gravity imply that the gauge-theory FFs satisfy hidden relations shown schematically as
\begin{equation}
\label{eq:generalfactorization}
\vec{v} \cdot \vec{\mathcal{F}}_n  \big|_{s_{\rm sp}=0}= \mathcal{F}_m \times \mathcal{A}_{n+2-m},
\end{equation}
where the special kinematics of spurious pole $s_{\rm sp}=0$ is considered.  
The  $\vec{v}$ vectors are rational functions of Mandelstams similar to the BCJ vectors in the BCJ relations for amplitudes \cite{Bern:2008qj}, and \eqref{eq:generalfactorization} may be also understood as generalized BCJ relations for FFs.

In this letter, we explain these properties using tree-level FFs $\mathcal{F}_{\operatorname{tr}(\phi^2)}$ in the Yang-Mills-scalar (YMS) theory.
Similar mechanism also applies to a much wider range of FFs, such as $\mathcal{F}_{\bar\psi\psi}$ in QCD which are equivalent to a class of Higgs amplitudes. Therefore, our discussion also provides for the first time a double copy for amplitudes involving a color singlet particle.
We will discuss various generalizations at the end of the paper.

\section{Invitation: a three-point example}
\begin{figure}[t]
    \centering
 \includegraphics[height=0.18\linewidth]{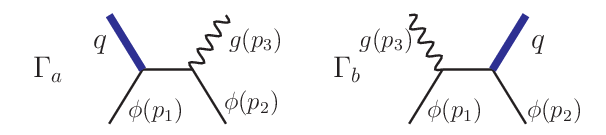}
    \caption{Feynman diagrams for the three-point gauge-theory FF. The blue thick line represents operator insertion carrying total momentum $q$. The black straight line and the wiggle line are  scalars and gluons respectively.  }
    \label{fig:F3tree}
\end{figure}

\noindent
The new features of the form-factor double copy can be mostly illustrated by considering the simple three-point FF 
$\itbf{F}_{3}(1^\phi, 2^\phi, 3^g)$ in YMS 
(see details of the theory in Supplemental Materials).
In this example, there are two cubic Feynman diagrams $\Gamma_{a,b}$ as given in Figure \ref{fig:F3tree}, and the full-color FF can be written as 
\begin{equation}\label{eq:3pttreefullcolor}
	\itbf{F} _{3}(1^{\phi},2^{\phi},3^{g})=\frac{C_{a}N_{a}(\varepsilon_3,\{p_{i}\})}{s_{23}}+\frac{C_{b}N_{b}(\varepsilon_3,\{p_{i}\})}{s_{13}}\,,
\end{equation}
where $C_{a,b}$ are color factors, and $N_{a,b}$ are kinematic numerators depending on  momenta $p_i, i=1,2,3$ and the gluon polarization vector $\varepsilon_3$.
We emphasize that the operator only couples to the scalar lines and $C_a{=}C_{b}{=}f^{123}$ since the color factor of the operator is a $\delta$-function in color space. 

To obtain the double copy, one can square the kinematic numerators and propose the following quantity in gravity:
\begin{equation}\label{eq:3pttreeGRA}
\begin{aligned}
	 \mathcal{G} _{3}(1^{\phi},2^{\phi},3^{h})
	&=\frac{N_{a}^2(\varepsilon_3,\{p_{i}\})}{s_{23}}+\frac{N_{b}^2(\varepsilon_3,\{p_{i}\})}{s_{13}}\,.
\end{aligned}
\end{equation}
If $\mathcal{G}_{3}$ is a well-defined quantity in gravity, it must be invariant under the diffeomorphism transformation, which acts on the graviton polarization tensor as $\varepsilon_{3}^{\mu\nu}{ \rightarrow }\varepsilon_{3}^{\mu\nu} {+} p_3^{(\mu} \xi^{\nu)}$. Here $\varepsilon_{3}^{\mu\nu}{=}\varepsilon_3^{(\mu}\varepsilon_3^{\nu)}$  with the brackets indicating the symmetric-traceless part and  and $\xi$ is a reference vector satisfying $\xi\cdot p_3{=}0$. 
However, if naively plugging in the numerators from Feynman rules for $N_{a,b}$, one easily finds that the diffeomorphism invariance is broken.

The key condition to restore the invariance is the color-kinematics duality, as in the amplitude cases \cite{Bern:2008qj, Bern:2010ue}.
Given the aforementioned color relation $C_a {=} C_b$, one can require the numerators to satisfy a parallel relation, defined as \emph{operator-induced dual relations}, and get
\begin{equation}\label{eq:3ptnumsols}
	N_{a}=N_{b}=\frac{s_{13}s_{23}}{s_{13}+s_{23}}\mathcal{F} _{3}(1^{\phi},3^{g},2^{\phi})\,,
\end{equation}
where $\mathcal{F} _{3}$ is the color-ordered three-point FF. 
Note that the CK-dual numerators are uniquely determined and manifestly gauge invariant. 
With this solution, the ${\cal G}_3$ proposed in \eqref{eq:3pttreeGRA} is given as 
\begin{equation}\label{eq:3pttreeGRAfinal}
	\mathcal{G}_3 =\frac{s_{13}s_{23}}{s_{13}+s_{23}}\left(\mathcal{F} _{3}(1^{\phi},3^{g},2^{\phi})\right)^2 ,
\end{equation}
which is indeed diffeomorphism invariant \footnote{To get \eqref{eq:3pttreeGRAfinal} it is enough to require only one copy of the numerators in \eqref{eq:3pttreeGRA} to satisfy \eqref{eq:3ptnumsols}, similar to the amplitude case \cite{Bern:2010yg}.}.

Getting ${\cal G}_3$ diffeomorphism invariant is not a free lunch though. 
In particular, the numerators \eqref{eq:3ptnumsols} contain a spurious pole $s_{13}{+}s_{23}$. In the gauge-theory FF, this pole is cancelled by summing up the two terms in \eqref{eq:3pttreefullcolor}; however, after double copy it becomes a real pole in \eqref{eq:3pttreeGRAfinal}.
Does this pole have a physical meaning in gravity?

Remarkably, the pole $s_{13}{+}s_{23}={-}(s_{12}{-}q^2)$ looks like a massive Feynman propagator, 
and the residue of ${\cal G}_3$ on this ``spurious'' pole can be nicely organized as 
\begin{equation}
\label{eq:G3factorization}
\mathrm{Res}\left[\mathcal{G}_3 \right]_{s_{12}=q^2}=\left(\mathcal{F} _{2}(1^{\phi},2^{\phi})\right)^2\times (\mathcal{A} _{3}(\QQ_2^{S}, 3^g, -q^{S}))^2 ,
\end{equation}
where $\mathcal{F} _{2}(1^{\phi},2^{\phi})=1$ is the minimal FF, and 
\begin{equation}
\mathcal{A} _{3}(\QQ_2^{S}, 3^g, -q^{S})=2\epsilon_3\cdot q \,, \quad \text{with }\QQ_2=p_1+p_2, \,
\end{equation}
is the three-point planar amplitude of a gluon and one pair of massive scalars $S$ with mass $m^2{=}q^2{=}\QQ_2^2$, see \emph{e.g.}~\cite{Badger:2005zh}.
In this sense, \eqref{eq:G3factorization} can be interpreted as a factorization formula
\begin{equation}\label{eq:3pttreeGRA2}
	\mathrm{Res}\left[\mathcal{G}_3 \right]_{s_{12}=q^2}=\mathcal{G}_2 (1^\phi,2^\phi) \  \mathcal{M}_{3}(\QQ_2^{S}, -q^{S}, 3^h) \,,
\end{equation} 
where $\mathcal{G}_2 {=}\big(\mathcal{F} _{2}\big)^2$ is the double copy of the minimal FF, and $ \mathcal{M}_{3}(\QQ_2^{S}, {-}q^{S}, 3^h){=} \big(\mathcal{A} _{3}(\QQ_2^{S}, 3^{g}, {-}q^{S})\big)^2$  is the three-point amplitude of a graviton coupled to two massive scalars (such amplitudes appear extensively in the gravitational wave studies via double copy, see \emph{e.g.}~\cite{Bern:2019crd}).

\begin{figure}[t]
    \centering
 \includegraphics[width=1\linewidth]{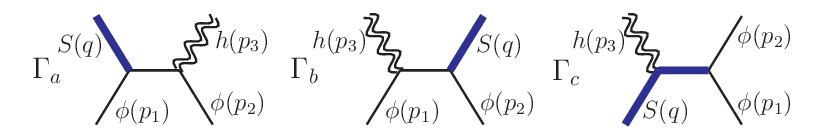}
    \caption{Feynman diagrams in gravity for the double copy of the three-point FF. The thick blue line in this case is the massive scalar with mass $m^2=q^2$. 
    The doubled wiggle line represents a graviton.
    Note that $\Gamma_c$ is a new diagram required by the factorization property \eqref{eq:3pttreeGRA2} on the new pole.}
    \label{fig:F3treeGRA}
\end{figure}

The factorization property \eqref{eq:3pttreeGRA2} implies that the spurious pole $s_{13}+s_{23}$ in gauge theory should be understood as a physical pole in gravity! 
Indeed, it represents the factorization of the Feynman diagram $ \Gamma_c$ in Figure \ref{fig:F3treeGRA}. 
The appearance of this new diagram is natural since gravitons couple to everything including the ``operator'' leg.

Furthermore, one can check that $\mathcal{G}_3 $ matches the expression from summing up all the three Feynman diagrams of Figure \ref{fig:F3treeGRA} in gravity: $\Gamma_a$ and $\Gamma_b$ contribute to the $s_{13}$ and $s_{23}$ poles while  the $s_{13}{+}s_{23}$ pole is from $\Gamma_c$.

In  addition, by taking the ``square-root'' of the double copy factorization \eqref{eq:G3factorization},  one can get an intriguing relation for the gauge-theory FF as the simplest example for \eqref{eq:generalfactorization}:
\begin{equation}
s_{13}\mathcal{F}_3(1^{\phi},3^{g},2^{\phi})  \big|_{s_{12}=q^2}= \mathcal{F}_2(1^{\phi},2^{\phi}) \  \mathcal{A} _{3}(\QQ_2^{S}, 3^g, -q^{S}) \,.
\end{equation}

We can summarize the main features in the three-point example as:
\begin{enumerate}
\item 
Diffeomorphism invariance requires that the numerators satisfy complete CK-dual relations, in particular those induced by the operator insertion; 

\item 
Such numerators contain poles which are spurious poles in gauge theory but can be interpreted as physical poles in gravity, on which the double-copy result factorizes as a product of double copies of the lower-point FF and amplitude; 

\item 
The gravity factorization implies an interesting relation for the gauge-theory FF.
While the three-point example seems too simple, we show below that the same features hold for higher-point cases.
\end{enumerate}

\section{General double copy }

\noindent
To generalize the above discussion to higher points, we develop a systematic way to obtain the CK-dual numerators and the KLT-type double copy. 
We consider the four-point FF $\itbf{F}_{4}(1^\phi, 2^\phi, 3^g, 4^g)$ as an explicit example,
  and the generalization to $n$-point is straightforward. 

\begin{figure}[t]
    \centering
 \includegraphics[width=0.95\linewidth]{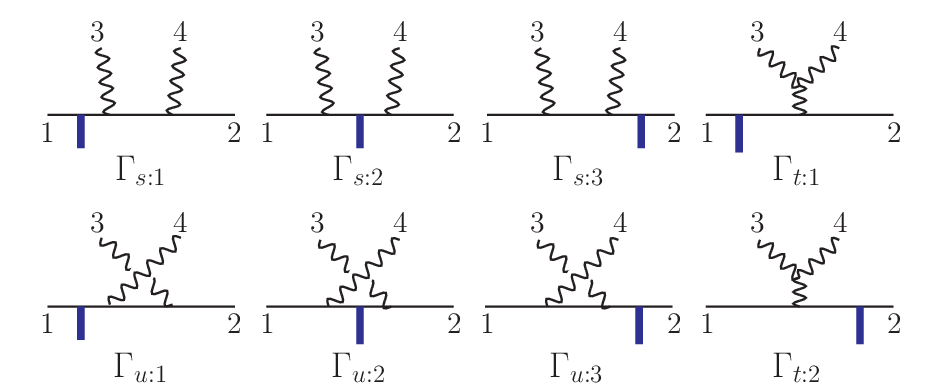}
    \caption{Cubic diagrams for the four-point FF in  $\itbf{F} _{4}$. }
    \label{fig:F4tree}
\end{figure}

As in the three-point case, we first express the four-point FF 
in terms of eight cubic diagrams shown in Figure \ref{fig:F4tree}:
\begin{equation}\label{eq:4pttreefullcolor}
\itbf{F} _{4}(1^{\phi},2^{\phi},3^{g},4^{g})= \sum_{\Gamma_i} \frac{C_{i}N_{i}}{D(\Gamma_i)}\,,
\end{equation}
where $D(\Gamma_i)$ is the propagators of $\Gamma_i$.
Since the color factor of the operator is a $\delta$-function, the eight $C_i$s take only three different values, and we can classify the diagrams into three groups accordingly.
For instance, the first three graphs $\Gamma_{s:i}, i{=}1,2,3$ share the same color factor $C_{s} {=} f^{a_1a_3b}f^{ba_4a_2}$; and $\Gamma_{t:i}$ and $\Gamma_{u:i}$ have color factors $C_t$ and $C_u$ respectively.
The three color factors satisfy the Jacobi relation $C_{s}{=}C_{t}{+}C_{u}$ like the four-point amplitude.

For the purpose of double copy, the CK duality is required, which asks the numerators to satisfy
\begin{align}
& N_{s:1}=N_{s:2}=N_{s:3} = N_{s} \,, \ \ \ \   N_{t:1}=N_{t:2} = N_{t} \,, \nonumber\\
& N_{u:1}=N_{u:2}=N_{u:3} = N_{u}\,, \ \ \ N_{s}=N_{t}+N_{u} \,.
\end{align}
Consequently, the FF can be written as
\begin{equation}
\label{eq:4pttreefullcolor-CK}
\itbf{F} _{4}= {C_{s} N_{s} \over P_{s}} +{C_{t}N_{t} \over P_{t}}+{C_{u} N_{u} \over P_{u}} \,,
\end{equation}
with $P_{s}^{-1}\equiv \sum_{i=1}^{3}(D(\Gamma_{s:i}))^{-1}$, and $P_{{t},{u}}$ are defined likewise. 
Alternatively, since $C_{s}, C_{u}$ form a color basis, $\itbf{F} _{4}$ can be expanded using color-ordered FFs $\mathcal{F}_{4}$ as
\begin{equation}
\label{eq:4pttreefullcolor-DDM}
\itbf{F} _{4}= C_{s} \mathcal{F}_{4} (1, 3,4,2) + C_{u} \mathcal{F}_{4} (1,4,3,2) \,,
\end{equation}
which is the Del Duca-Dixon-Maltoni (DDM) color decomposition \cite{DelDuca:1999rs}. 

Matching \eqref{eq:4pttreefullcolor-DDM} with \eqref{eq:4pttreefullcolor-CK}, one finds the following relation:
\begin{equation}\label{eq:4ptFNeq0}
\vec{\mathcal{F}}_4 = \Theta^{\mathcal{F}}_{4} \cdot \vec N_4\,, \ 
\vec{\mathcal{F}}_4 =
\begin{pmatrix}
\mathcal{F}_{4} (1, 3,4, 2) \\
\mathcal{F}_{4} (1, 4,3, 2)
\end{pmatrix} ,
\ \vec N_4 =
\begin{pmatrix}
N_s \\
N_u
\end{pmatrix} ,
\end{equation}
where  $\Theta^{\mathcal{F}}_{4}$ is a matrix of propagators as
\begin{equation}
\label{eq:4ptFNeq1}
\Theta^{\mathcal{F}}_{4} = 
\begin{pmatrix}
{1\over P_{s}} +  {1\over P_{t}}& - {1\over P_{t}} \\
-{1\over P_{t}} &  {1\over P_{t}} + {1\over P_{u}}
\end{pmatrix} .
\end{equation}
One can check that $\Theta^{\mathcal{F}}_{4}$ has full rank \footnote{This is different from the amplitude case. For $n$-point amplitudes,  a similarly defined $(n-2)!\times(n-2)!$ matrix $\Theta_n$ has rank $(n-3)!$, which implies that there are $(n-3)(n-3)!$ BCJ relations among the planar amplitudes.} and thus by simply inverting \eqref{eq:4ptFNeq1} one obtains the CK-dual numerators as 
\begin{equation}\label{eq:4ptnumsols}
	N_4[\alpha]=\sum_{\beta\in S_2} 
	{\bf S}^{{\cal F}}_{4}[\alpha|\beta] \, \mathcal{F}_{4} [\beta] , \qquad {\bf S}^{{\cal F}}_{4}\equiv\left(\Theta^{\mathcal{F}}_{4}\right)^{-1} ,
\end{equation} 
where $\alpha,\beta$ label the vector/matrix components in \eqref{eq:4ptFNeq0}.
These CK-dual numerators contain spurious poles (such as \eqref{eq:4ptNsSolution}) which are introduced by ${\bf S}^{{\cal F}}_{4}$, and we will discuss more on this shortly.

Given the CK-dual numerators \eqref{eq:4ptnumsols}, one can perform the double copy to get the gravitational quantity as
\footnote{As in the three-point example, here only one copy of the numerators are required to satisfy CK-dual property in getting \eqref{eq:4pttreeGRA} and the other copy can be chosen as from Feynman rules; in thus case, the double-copy in general is given as a sum of eight terms corresponding to eight trivalent diagrams in Figure \ref{fig:F4tree}.}
\begin{equation}\label{eq:4pttreeGRA}
	\mathcal{G} _{4}={N_{s}^2 \over P_{s}}\hskip -1pt+\hskip -1pt{N_{t}^2 \over P_{t}}\hskip -1pt+\hskip -1pt{N_{u}^2 \over P_{u}}=\hskip -4pt \sum_{\alpha,\beta\in S_2}\hskip -4pt \mathcal{F}_{4} [\alpha] {\bf S}^{\mathcal{F}}_{4}[\alpha|\beta] \mathcal{F}_{4} [\beta],
\end{equation}
which is manifestly diffeomorphism invariant. 
Such a bilinear form is very similar to the KLT form for amplitudes, 
and ${\bf S}^{{\cal F}}_{4}$ serves as the (four-point) KLT kernel.

Clearly, \eqref{eq:4pttreeGRA} can be easily generalized to $n$-points (2 scalars plus $n{-}2$ gluons) such that the double copy in the KLT form is
\begin{equation}\label{eq:genericKLT}
\mathcal{G}_{n} =\sum_{\alpha,\beta\in S_{n-2}}\mathcal{F} _{n}[\alpha] \, \mathbf{S}^{\mathcal{F}}_{n}[\alpha|\beta] \, \mathcal{F} _{n}[\beta]\,,
\end{equation}
where ${\bf S}^{{\cal F}}_{n}\equiv(\Theta^{\mathcal{F}}_n)^{-1}$ is the $n$-point KLT kernel determined by the propagator matrix $\Theta^{\mathcal{F}}_n$ and $\mathcal{F} _{n}[\alpha]$ are color-ordered FFs in the DDM basis. 

We have verified that the double copy construction is consistent with physical requirements by considering all factorization channels. 
The factorizations on ``physical" poles (appearing in the gauge FFs) are similar to ordinary amplitudes, and the main concern here is about the new ``spurious''-type poles. 
Below we show that they indeed become  physical propagators on which ${\cal G}_n$ has nice factorization properties, as in the three-point case.

\section{``Spurious'' poles and factorizations}

\noindent
The sign of the ``spurious'' poles first appear in the determinant of  the propagator matrices $\Theta^{{\cal F}}$. 
In the four-point example above, one finds 
\begin{equation}
\label{eq:detTheta4}
\det ( \Theta^{{\cal F}}_{4}) = {(q^2 - s_{12})(q^2 - s_{123})(q^2 - s_{124}) \over s_{13} s_{14} s_{23} s_{24} s_{34} s_{134} s_{234} }\,.
\end{equation}
Importantly, the numerators of $\det( \Theta^{{\cal F}}_{4})$ is a product of ``spurious'' poles such as $ s_{1 2 3}{-} q^2$. 
Similar structures extend to higher-point cases, where the matrix $\Theta^{{\cal F}}_{n}$ can be rather complicated but its determinant remains strikingly simple, and its numerator are composed of  ``spurious'' poles. 

The information of poles of the KLT kernel ${\bf S}^{{\cal F}}$ is provided by the zeros of $\det (\Theta^{{\cal F}})$. 
Specified to ${\bf S}^{{\cal F}}_{4}$, there are only simple poles 
 $\{s_{12}{-}q^2,  s_{123}{-}q^2, s_{124}{-}q^2\}$ like massive Feynman propagators
\footnote{We also remark that none of the propagators (physical poles) such as $s_{13}$ appears in the denominators of matrix elements of  $ {\bf S}^{{\cal F}}_{4}$, so that the master numerators also have no such poles. This is similar to the amplitude case where the master numerators can be select to be local, which further explains the physical pole factorization of the double copy.}.
For higher $n$-point cases, we have explicitly checked that $ {\bf S}^{{\cal F}}_{n}$ also exhibits only simple poles like $(s_{(12\cdots)}{-}q^2)$, where ${\cdots}$ represents gluon momenta, up to highly non-trivial seven points.

Furthermore, the numerators $N_4$ also  have only the spurious-type simple poles. For example, $N_{s}$ in \eqref{eq:4ptFNeq0} can be given as 
\begin{align}
	N_{s}=&-\frac{2\left(f_{3}^{\mu\nu}f_{4,\nu\rho}p_{1,\mu}p_{2}^{\rho}\right)}{(s_{12}-q^2)}+ 
	\frac{4\left(f_{3}^{\mu\nu}p_{1,\mu}p_{2,\nu}\right)\left(f_{4}^{\mu\nu}p_{2,\mu}q_{\nu}\right)}{(s_{12}-q^2)(s_{123}-q^2)}\nonumber \\
	&+\frac{4\left(f_{4}^{\mu\nu}p_{1,\mu}p_{2,\nu}\right)\left(f_{3}^{\mu\nu}p_{1,\mu}q_{\nu}\right)}{(s_{12}-q^2)(s_{124}-q^2)}\,,
	\label{eq:4ptNsSolution}
\end{align}
where $f^{\mu\nu}_{i}{=} \varepsilon_i^{\mu}p_i^{\nu}{-}\varepsilon_i^{\nu}p_i^{\mu}$. 
Consequently, the pole structure of ${\bf S}^{{\cal F}}$ and $N$  guarantees that the double copy $\mathcal{G}$ contains the spurious-type \emph{simple} poles. 

\begin{figure}[t]
    \centering
 \includegraphics[width=1\linewidth]{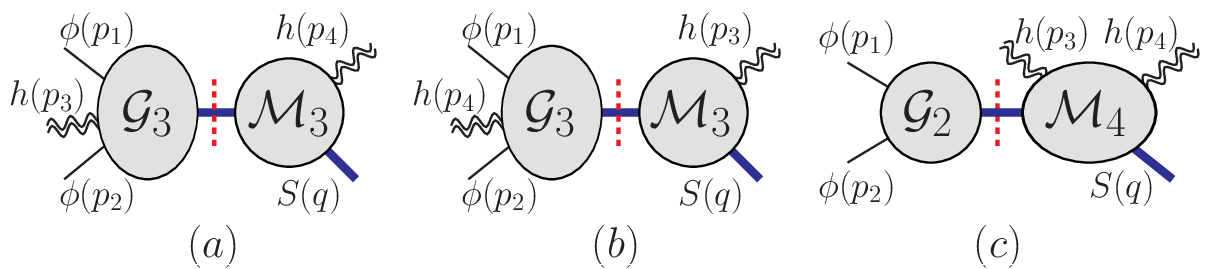}
    \caption{Factorization of $\mathcal{G}_4$ on the three new poles.} 
    \label{fig:4ptfactorizationGRA}
\end{figure}

It is necessary to check such new poles are well-defined physical poles in $\mathcal{G}_{n} $.
For ${\cal G}_4$, the factorizations on the three new poles are presented in Figure \ref{fig:4ptfactorizationGRA}, \emph{e.g.}, on the pole $ s_{123}{-} q^2$ it can be written as
\begin{equation}\label{eq:4ptfactorizationGRA}
\begin{aligned}
	\mathrm{Res}\left[\mathcal{G}_4 \right]_{s_{123}=q^2}=&\mathcal{G}_3 (1,2,3)\  \mathcal{M}_{3}(\QQ_3^{S}, -q^{S},4^h) \,,
\end{aligned}
\end{equation}
where $\QQ_3{=}p_1{+}p_2{+}p_3$. 
As in the three-point case, these factorization relations show that new Feynman diagrams (with massive propagators) contribute. 

For the generic $n$-point cases, we have 
\begin{equation}\label{eq:nptfactorizationGRA}
\mathrm{Res}\left[\mathcal{G}_{n} \right]_{q^2_{m}=q^2}=\mathcal{G}_{m} (1,..,m)\ \mathcal{M}_{m'}(\QQ_{m}^{S}, -q^{S},{m+1},..,n)\,,
\end{equation}
where $\QQ_{m}=\sum_{i=1}^{m}p_i$, $m'=n{+}2{-}m$ and $\mathcal{M}_{m'}$ is an $m'$-point amplitude of gravitons coupled to a pair of massive scalars. 
We have checked this up to seven points.

\section{Hidden relation of gauge-theory FFs}

\noindent
The double-copy factorization relations \eqref{eq:nptfactorizationGRA} imply new relations for gauge FFs.
For instance,  \eqref{eq:4ptfactorizationGRA} implies the following relation for $\mathcal{F}_4$
\begin{equation}
\label{eq:4ptbcjB}
(\vec{v}_4 \cdot \vec{\mathcal{F}}_{4})\big|_{s_{123}=q^2}
 =\mathcal{F}_{3}(1^{\phi},3^{g},2^{\phi}) \ \mathcal{A}_{3}(\QQ_{3}^{S},4^{g},-q^{S}) \,,
\end{equation}
where the (row) vector $\vec{v}_4$ and (column) vector $\vec{\mathcal{F}}_4 $ are
\begin{align}
&\vec{v}_4 =\left(\tau_{42}, \tau_{42} + \tau_{43}\right)\,, \quad
\vec{\mathcal{F}}_4 =
\begin{pmatrix}
\mathcal{F}_{4} (1, 3,4, 2) \\
\mathcal{F}_{4} (1, 4,3, 2)
\end{pmatrix} \,,
\end{align}
with $\tau_{ij}{=}2 p_i \cdot p_j$ \footnote{Here we use $\tau_{ij}=2p_i\cdot p_j$ instead of $s_{ij}=(p_i+p_j)^2$ because it is easier to generalize to  higher points. Besides, the $\tau$ notation is also valid for the massive $\phi$ case.}.
One may notice that this is reminiscent of the BCJ relation for four-point amplitudes \cite{Bern:2008qj}:
\begin{equation}
\tau_{42} \mathcal{A}_{4}(1,3,4,2) + (\tau_{42}+\tau_{43}) \mathcal{A}_{4}(1,4,3,2) = 0 \,.
\end{equation}
Here the RHS of \eqref{eq:4ptbcjB} is not zero; instead, it offers a relation involving FFs with different numbers of external legs and scalar-Yang-Mills amplitudes. 

Similar relations exist for higher-point cases taking the schematic form as \eqref{eq:generalfactorization}. Such a general $n$-point relation is
\begin{align}
\label{eq:nptbcjN}
&\sum_{i=3}^{n}\tau_{n,(2+i+ .. + (n-1))}\mathcal{F}_{n}{\small (1, 3,.., i-1,n,i, .. ,n-1,2)} \big|_{\QQ_{n-1}^2=q^2} \nonumber\\
& \hskip -2pt = \mathcal{F}_{n-1}(1,3, .. , n-1,2) \, \mathcal{A}_3(\QQ_{n-1},n,-q),
\end{align}
where $\QQ_{n-1}\ {=}\sum_{i=1}^{n-1}p_i$ and $\tau_{n,(j+..+k)}{=} 2 p_n \cdot (p_j+..+p_k)$. 
A proof for this relation for MHV FFs and complete relations for four and five-point FFs are given in Supplementary Materials. 

The bridge between the gauge theory relation \eqref{eq:4ptbcjB} and the gravity factorization \eqref{eq:4ptfactorizationGRA} is a decomposition relation of ${\bf S}^{\mathcal{F}}_{4}$, in which $\vec{v}_4$ also plays a central role. 
Concretely, for the $2\times 2$ matrix ${\bf S}^{\mathcal{F}}_{4}$, we have
\begin{equation}
\label{eq:4ptSfdecomp}
\mathrm{Res}\left[{\bf S}^{\mathcal{F}}_{4}\right]_{s_{123}=q^2}=\vec{v}_4^{\rm \scriptscriptstyle T} \cdot ({\bf S}^{\mathcal{F}}_{3}\otimes {\bf S}^{\mathcal{A}}_{3}) \cdot \vec{v}_4\,,
\end{equation}
where ${\bf S}^{\mathcal{F}}_{3}$ is a $1\times 1$ matrix $\big(\frac{s_{13}s_{23}}{s_{13}{+}s_{23}}\big)$ in \eqref{eq:3pttreeGRAfinal} and ${\bf S}^{\mathcal{A}}_{3}=\big(1\big)$ is the KLT kernel for three-point amplitudes. Thus, we can derive \eqref{eq:4ptfactorizationGRA} from \eqref{eq:4ptbcjB}
\begin{align}
\label{eq:4ptFactorization}
\mathrm{Res}&\left[\mathcal{G}_{4} \right]_{\QQ_3^2=q^2} 
= \mathcal{F} _{4} \cdot \mathrm{Res}\left[\mathbf{S}^{\mathcal{F}}_{4} \right] \cdot \mathcal{F} _{4}  \big|_{\QQ_3^2=q^2} \\
&\quad\ \ =(\vec{v}_4\cdot \mathcal{F}_4) (\mathbf{S}^{\mathcal{F}}_{3} \otimes \mathbf{S}^{\mathcal{A}}_{3}) (\vec{v}_4\cdot \mathcal{F}_4) \big|_{\QQ_3^2=q^2} \nonumber \\
&\quad\ \ =  (\mathcal{F}_3 {\cal A}_{3})  (\mathbf{S}^{\mathcal{F}}_{3} \otimes \mathbf{S}^{\mathcal{A}}_{3})  (\mathcal{F}_3 {\cal A}_{3}) \nonumber\\
&\quad\ \ = (\mathcal{F} _{3} \cdot \mathbf{S}^{\mathcal{F}}_{3} \cdot \mathcal{F} _{3}) (\mathcal{A} _{3} \cdot \mathbf{S}^{\mathcal{A}}_{3} \cdot \mathcal{A} _{3}) = \mathcal{G}_{3} \mathcal{M}_{3}. \nonumber
\end{align}
For the generic $n$-point case, one has the similar decomposition as \eqref{eq:4ptSfdecomp}, reading
\begin{equation}
\label{eq:NptSfdecomp}
\mathrm{Res}\left[{\bf S}^{\mathcal{F}}_{n}\right]_{\QQ_m^2=q^2}= {\bf V}_n^{\rm \scriptscriptstyle T} \cdot ({\bf S}^{\mathcal{F}}_{m} \otimes {\bf S}^{\mathcal{A}}_{m^{\prime}}) \cdot {\bf V}_n \,,
\end{equation}
where the matrix ${\bf S}^{\mathcal{F}}_{n}$ is not full-ranked when taking the residue
and can be factorized into
${\bf S}^{\mathcal{F}}_{m} \otimes {\bf S}^{\mathcal{A}}_{m^{\prime}}$ through the (rectangular) matrix ${\bf V}_n$. Here ${\bf V}_n$  is a matrix as a collection of BCJ-like vectors $\vec{v}$ (like $\vec{v}_4$); and interestingly, such $\vec{v}$ vectors are exactly those appearing in the hidden factorization relation \eqref{eq:generalfactorization}.
As a direct consequence, similar to \eqref{eq:4ptFactorization}, \eqref{eq:nptfactorizationGRA} can be derived assuming \eqref{eq:generalfactorization} and \eqref{eq:NptSfdecomp}. We explain these relations in more detail in Supplementary Materials.

\section{Discussion}

\noindent
A wide range of generalizations for the above FF double copy can be realized:
\begin{enumerate}
\item 
An important generalization is for the double copy of amplitudes involving a color singlet particle.
One notable case is the FFs of a bilinear quark operator $\bar\psi\psi$, which are equivalent to the tree-level amplitudes ${\cal A}(q^H, 1^{\bar\psi}, 2^\psi, 3^g, \ldots, n^g)$ in QCD involving a color singlet Higgs particle ($m_H^2{=}q^2$).
The corresponding double-copy quantities are the amplitudes of Higgs plus two photons and an arbitrary number of gravitons. See Supplementary Materials for further details. 

\item 
Double copy for FFs with more than two external scalar lines are also achieved. These FFs can involve either high-length operators ${\rm tr}(\phi^L)$ or the ${\rm tr}(\phi^2)$ operator but with bi-adjoint-scalar interactions in the theory. 
For instance, for three scalars, the double copy of $\mathcal{F}_{{\rm tr}(\phi^3)}$ and $\mathcal{F}_{{\rm tr}(\phi^2)}$ with three scalar states can be obtained which involve simple spurious poles like $(s_{123\cdots}{-}q^2)$.

\item 
A double-copy prescription for the FFs of  $\operatorname{tr}(F^2)$ with pure gluons is possible and engages a gauge invariant expansion and a mixture of different BCJ numerators appearing in this expansion.

\item 
The CK-dual numerators, like \eqref{eq:4ptNsSolution}, have interesting structures reminiscent of the HEFT BCJ numerators constructed from kinematic Hopf algebra \cite{Brandhuber:2021kpo, Brandhuber:2021bsf}. Such a connection to the Hopf algebra leads to an all-multiplicity closed formula for the master numerators, see \cite{Chen:2022nei}.   

\end{enumerate}

\noindent All these generalizations will be discussed in detail in \cite{treepaper1, treepaper2}. 

We mention some other problems associated with the observations made in this paper:
\begin{itemize}
\item 
An important natural question is to apply the above tree-level picture to the loop level.
In particular, the previous studies of  CK-dual FFs \cite{Boels:2012ew,Yang:2016ear, Lin:2020dyj, Lin:2021kht,Lin:2021qol} considered only Jacobi relations, and further understanding about the operator-induced relations will be crucial for the loop-level double copy.
\item 
Our finding suggests that a more general class of KLT matrices containing spurious poles can be physically meaningful, and it would be interesting to explore this using the KLT bootstrap method \cite{Chi:2021mio}. 
\item 
Our double copy is valid in $D$-dimension, and it is interesting to pursue a CHY formula \cite{Cachazo:2013hca, Cachazo:2014xea}(and the corresponding double copy) for FFs, generalizing the previous 4-dimensional connected description of FFs \cite{He:2016jdg, Brandhuber:2016xue}.
In this direction, noting that the propagator matrix $\Theta^{\cal F}$ can be regarded as the FF in bi-adjoint scalar theory \cite{Du:2011js, Bjerrum-Bohr:2012kaa, Cachazo:2013iea} could be helpful. See some recent progress in \cite{Dong:2022bta}.
\item 
As in the original KLT prescription \cite{Kawai:1985xq}, it would be very interesting to have a string theory generalization for the KLT-like formula \eqref{eq:genericKLT} and also find a $Z$-function \cite{Broedel:2013tta, Carrasco:2016ldy} that encodes the $\alpha'$-expansion for FFs. 
\item 
Finally, conjectured closed formulae for the BCJ-like relations \eqref{eq:generalfactorization} of FFs will be given in \cite{treepaper2}. Having a better understanding and proof, using the field theory \cite{Feng:2010my} or in particular the string theory methods \cite{Bjerrum-Bohr:2009ulz, Stieberger:2009hq}, should be fascinating.

\end{itemize}

\vskip .3cm
{\it Acknowledgments.}
We thank Gang Chen, Jin Dong, Song He, Congkao Wen, Mao Zeng and especially Henrik Johansson and Radu Roiban for enlightening discussions. 
This work is supported by the National Natural Science Foundation of China (Grants No.~11822508, 11935013, 12047502, 12047503, 11947301), and by the Key Research Program of the Chinese Academy of Sciences, Grant NO. XDPB15.

\bibliographystyle{apsrev4-2}

\bibliography{DoubleCopyLetter}


\newpage

\appendix

\onecolumngrid
\section{Supplemental material}

\subsection{I. Some details of the scalar-YM theory and the Higgs theory}

For reader's convenience, we provide some details about the theories we mentioned in the main text.

\subsubsection*{Scalar-YM theory}

The action of the scalar-Yang-Mills theory can be chosen as
\begin{equation}
{\cal L}^{\phi{\rm YM}} = -{1\over2}{\rm tr}(F_{\mu\nu}F^{\mu\nu}) +  {1\over2}{\rm tr}(D^\mu \phi D_\mu \phi) \,. 
\end{equation}
The gauge field $A_\mu = A_\mu^a T^a$ and the scalar $\phi = \phi^a T^a$ are both in the adjoint representation, where $T^a$ are the generators of gauge group satisfying $[T^a, T^b] = i f^{abc} T^c$.
The covariant derivative acts as
$D_\mu \  \cdot = \partial_\mu \cdot + i g [A_\mu, \cdot \ ]$, and $[D_\mu, D_\nu] \ \cdot = i g [F_{\mu\nu}, \cdot \ ]$.
An $n$-point form factor of the operator ${\rm tr}(\phi^2)$ is defined as the following matrix element:
\begin{equation}
\itbf{F}_n(1^{\phi},2^{\phi},3^{g}, \ldots, n^g) =  \int d^{D} x \, e^{-i q \cdot x}\langle \phi(p_1) \, \phi(p_2) \, g(p_3) \ldots g(p_n) |{\rm tr}(\phi^2)(x)| 0\rangle \,.
\end{equation} 
The two-point minimal form factor is simply
\begin{equation}
\itbf{F}_2(1^{\phi},2^{\phi}) = (2\pi)^4 \delta^{(4)}(q - p_1 - p_2) \delta^{a_1 a_2}  \,,
\end{equation}  
which has a trivial kinematic part equal to one, and one can make a double-copy directly. Thus the first interesting case is the three-point case as discussed in the main text.  The Feynman diagrams for the three-point case is clearly given as in Figure~\ref{fig:F3tree}.

The form factor can be understood as a tree-level amplitude in the the following theory
\begin{equation}
{\cal L}^{\phi{\textrm{YM-Higgs}}} = - {1\over2} {\rm tr}(F_{\mu\nu}F^{\mu\nu}) + {1\over2}  {\rm tr}(D^\mu \phi D_\mu \phi) + H {\rm tr}(\phi^2) + {1\over2} \partial^\mu H \partial_\mu H - {1\over2} m_H^2 H^2 \,,
\end{equation}
where $\phi = \phi^a T^a$ is a charged scalar while $H$ is a color-singlet scalar, and
\begin{equation}
\itbf{F}_{n} \Rightarrow \itbf{A}^{\phi{\textrm{YM-Higgs}}}_{n+1}(1^{\phi},2^{\phi},3^{g}, \ldots, n^g, q^H) =  \langle \phi(p_1) \, \phi(p_2) \, g(p_3) \ldots g(p_n) | H(q) \rangle \,.
\end{equation} 
Via double-copy, one has the gravitational theory
\begin{equation}
{\cal L}^{\phi{\textrm{Gra-Higgs}}} = \sqrt{-g} \Big({1\over16\pi G} R + {1\over2} g^{\mu\nu} \partial_\mu \phi \partial_\nu \phi + H \phi^2 + {1\over2} g^{\mu\nu} \partial_\mu H \partial_\nu H - {1\over2} m_H^2 H^2 \Big) \,,
\end{equation}
and the corresponding map is
\begin{equation}
A^{a}_\mu \rightarrow h_{\mu\nu}\,, \quad \phi^a \rightarrow \phi \,, \quad H \rightarrow H \,,
\end{equation}
where both $\phi$ and $H$ are scalars in the gravitational theory.
At tree-level, the double-copy of the form factor can be understood as the amplitude in the gravitational theory as
\begin{equation}
{\cal G}_n \Rightarrow M^{\phi{\textrm{Gra-Higgs}}}_{n+1}(1^{\phi},2^{\phi},3^{h}, \ldots, n^h, q^H) =  \langle \phi(p_1) \, \phi(p_2) \, h(p_3) \ldots h(p_n) | H(q) \rangle \,.
\end{equation}

\subsubsection*{Generalization to Standard Model Higgs amplitudes}

As mentioned in the discussion section, our double copy procedure can be applied to amplitudes involving the Standard Model Higgs particle. The related Lagrangian of the theory is given as
\begin{equation}
{\cal L}^{{\textrm{QCD-Higgs}}} = -{1\over2} {\rm tr}(F_{\mu\nu}F^{\mu\nu}) + i \bar{\psi} \gamma^\mu D_\mu \psi + H \bar{\psi} \psi + {1\over2} \partial^\mu H \partial_\mu H - {1\over2} m_H^2 H^2  \,,
\end{equation}
where $H$ is the color-singlet Higgs, and $\psi, \bar\psi$ are the quark and anti-quark in fundamental representation (here for simplicity we consider the flavor number to be one). 
The Higgs amplitudes that contain two external quarks and $n-2$ gluons are equivalent to the $n$-point form factor of the operator ${{\cal O}=\bar\psi\psi}$:
\begin{equation}
\itbf{A}(q^H, 1^{\bar\psi}, 2^{\psi}, 3^g, \ldots, n^g) = \itbf{F}_{\bar\psi\psi}(1^{\bar\psi}, 2^{\psi}, 3^g, \ldots, n^g) \,.
\end{equation}
Since there are two fundamental quarks, it should be clear that the trivalent diagrams have the same structure as that of  $\itbf{F}(1^{\phi},2^{\phi},3^{g},\ldots,n^{g})$.  Importantly, this implies that the propagator matrix $\Theta^{\mathcal{F}}_n$ of $\itbf{F}_{\bar\psi\psi}(1^{\bar\psi}, 2^{\psi}, 3^g, \ldots, n^g)$ is also exactly the same as the latter. Like \eqref{eq:genericKLT}, we have
\begin{equation}\label{eq:genericKLTferm}
		\mathcal{\tilde{G}}_{n} =\sum_{\alpha, \beta\in S_{n-2}}\mathcal{F} _{n}[1^{\bar{\psi}},\alpha,2^{\psi}] \mathbf{S}_{n}^{\mathcal{F}}[\alpha|\beta] \mathcal{F} _{n}[1^{\bar{\psi}},\beta,2^{\psi}]\,.
\end{equation}
Consider the three-point case as an explicit example. There are two trivalent diagrams similar to that in Figure~\ref{fig:F3tree}, which also share the same color factor.
The color-ordered three-point form factor can be obtained using \emph{e.g.}~Feynman diagrams as
\begin{equation}
    \mathcal{F}_{3}(1^{\bar{\psi}},3^{g},2^{\psi})=\frac{1}{2}\left(\frac{1}{s_{13}}+\frac{1}{s_{23}}\right)u_{1,\alpha} \left(\slashed{f_3}\right)^{\alpha}_{\beta} v_2^{\beta} + \frac{2 p_1\cdot f_3\cdot p_2}{s_{13}s_{23}} u_{1,\alpha} v_2^{\alpha}\,.
\end{equation}
The CK duality then requires the CK-dual numerator to be (same as \eqref{eq:3ptnumsols})
\begin{equation}
    N^{\cal F}_{3}[1^{\bar{\psi}},3^{g},2^{\psi}]=\frac{1}{2}u_{1,\alpha} \left(\slashed{f_3}\right)^{\alpha}_{\beta} v_2^{\beta} - \frac{2 p_1\cdot f_3\cdot p_2}{s_{12}-q^2} u_{1,\alpha} v_2^{\alpha}\,.
\end{equation}
The corresponding gravitational quantity via double copy is 
\begin{equation}
\mathcal{\tilde{G}}_3 =\frac{s_{13}s_{23}}{s_{13}+s_{23}}\left(\mathcal{F} _{3}(1^{\bar\psi},3^{g},2^{\psi})\right)^2\,.
\end{equation}
We have checked this for higher-point cases.
More details will be presented in another work.

Finally, we mention that the double-copy quantities at tree level can be understood as Higgs amplitudes 
$M(q^H, 1^g, 2^g, 3^h, \ldots, n^h)$
in the gravitational theory
\begin{equation}
{\cal L}^{{\textrm{Gra-Higgs}}} = \sqrt{-g} \Big({1\over16\pi G} R - {1\over4} F_{\mu\nu}F^{\mu\nu} + H F_{\mu\nu}F^{\mu\nu} + {1\over2} g^{\mu\nu} \partial_\mu H \partial_\nu H - {1\over2} m_H^2 H^2 \Big) \,.
\end{equation}
Here the double-copy map is
\begin{equation}
A^{a}_\mu \rightarrow h_{\mu\nu}\,, \quad \psi_i \rightarrow A^{\scriptscriptstyle{\rm U(1)}}_{\mu} \,, \quad H \rightarrow H \,,
\end{equation}
where the double copy of a (massless) fermion becomes a photon while the operator $H\mathcal{O}=H\bar{\psi}\psi$ representing the Yukawa coupling becomes now $H F^{\mu\nu}F_{\mu \nu}$. 

\subsection{II. New factorization relations for four- and five-point form factors}

\noindent
In this appendix, we give complete generalized BCJ vectors for four- and five-point form factors and the related identities. 
We use the notation $\QQ_{m}=\sum_{i=1}^{m}p_i$, $\tau_{ij}=2 p_i \cdot p_j$ and $\tau_{n,(j+..+k)}= 2 p_n \cdot (p_j+..+p_k)$.

\subsubsection{The four-point case} 

\noindent
The four-point case has two types of factorization relations, corresponding to (A) two-particle $s_{12}=q^2$ and (B) three-particle $s_{123}=q^2$ spurious poles (the result for the $s_{124}=q^2$ pole can be obtained by a trivial relabeling), respectively. Introducing the form factor basis $\vec{\mathcal{F}}_{4}$ as a (column) vector
\begin{equation*}
	\vec{\mathcal{F}}_{4}^{\rm \scriptscriptstyle T}=\left\{ \mathcal{F}_{4}(1,3,4,2),\mathcal{F}_{4}(1,4,3,2)\right\},
\end{equation*}
we have
\begin{align}
(\vec{v}^{A}_4 \cdot \vec{\mathcal{F}}_{4})\big|_{s_{12}=q^2} & =\mathcal{F}_{2}(1^{\phi},2^{\phi}) \  \mathcal{A}_{4}(\QQ_{2}^{S},3^{g},4^{g},-q^{S})\,, \\
(\vec{v}^{B}_4 \cdot \vec{\mathcal{F}}_{4})\big|_{s_{123}=q^2} & =\mathcal{F}_{3}(1^{\phi},3^{g},2^{\phi}) \ \mathcal{A}_{3}(\QQ_{3}^{S},4^{g},-q^{S}) ,
\end{align}
where
\begin{align}
&\vec{v}^{A}_4 =\Big\{ \frac{\tau_{31}\tau_{42}}{\tau_{3\QQ_2}}, \frac{ \tau_{32}\tau_{41}}{\tau_{3\QQ_2}} \Big\}\,,  \qquad 
\vec{v}^{B}_4 =\left\{\tau_{42}, \tau_{42} + \tau_{43}\right\} ,
\end{align}
with $\tau_{ij}=2 p_i \cdot p_j$.
The vectors $\vec{v}$ also appear in the decomposition of the KLT kernel as
\begin{align}
	 \mathrm{Res}\left[{\bf S}^{\mathcal{F}}_{4}\right]_{s_{12}=q^2} & =({\vec v}^{A}_4)^{\rm \scriptscriptstyle T} \ \left({\bf S}^{\mathcal{F}}_{2}\otimes{\bf S}^{\mathcal{A}}_{4}\right) \ {\vec v}^{A}_4\big|_{s_{12}=q^2} \,, \\
	 \mathrm{Res}\left[{\bf S}^{\mathcal{F}}_{4}\right]_{s_{123}=q^2} & =(\vec{v}^{B}_4)^{\rm \scriptscriptstyle T} \ ({\bf S}^{\mathcal{F}}_{3}\otimes {\bf S}^{\mathcal{A}}_{3}) \ \vec{v}^{B}_4 \big|_{s_{123}=q^2} \,.
\end{align}

Besides, we also give explict form of the four-point propagator matrix $\Theta_4^{\cal F}$ and the KLT kernel ${\bf S}_{4}^{\cal F}$:
\begin{align}
\Theta^{\mathcal{F}}_{4}& = 
\begin{pmatrix}
{1\over s_{13}s_{24}} +  {1\over s_{13}s_{134}} + {1\over s_{34}s_{134}} + {1\over s_{24}s_{234}} + {1\over s_{34}s_{234}} & - {1\over s_{34}s_{134}} - {1\over s_{34}s_{234}} \\
- {1\over s_{34}s_{134}} - {1\over s_{34}s_{234}} &  {1\over s_{23}s_{14}} +  {1\over s_{14}s_{134}} + {1\over s_{34}s_{134}} + {1\over s_{23}s_{234}} + {1\over s_{34}s_{234}}
\end{pmatrix} , \\
{\bf S}_{4}^{\cal F}& = 
\begin{pmatrix}
\frac{s_{13}s_{24}s_{34}(s_{134}s_{234}+s_{14}s_{134}+s_{23}s_{234})}{(s_{12}-q^2)(s_{123}-q^2)(s_{124}-q^2)}- \Delta_{t} & \Delta_{t}\\
 \Delta_{t}  &  \frac{s_{14}s_{23}s_{34}(s_{134}s_{234}+s_{24}s_{234}+s_{13}s_{134})}{(s_{12}-q^2)(s_{123}-q^2)(s_{124}-q^2)}- \Delta_{t}
\end{pmatrix}\,.
\end{align}
where $\Delta_{t}=  \frac{s_{13}s_{14}s_{23}s_{24}}{s_{12}-q^2}\left({1\over   s_{123}-q^2} + {1\over s_{124}-q^2}\right)$.

\subsubsection{The five-point case} 
\noindent
The five-point case has three types of generalized BCJ vectors, corresponding to the (A) $s_{12}=q^2$, (B) $s_{123}=q^2$ and (C) $s_{1234}=q^2$ poles respectively. 
We introduce the form factor basis $\vec{\mathcal{F}}_{5}$ as the following vector
\begin{equation*}
	\vec{\mathcal{F}}_{5}^{\rm \scriptscriptstyle T}=\left\{ \mathcal{F}_{5}(1,3,4,5,2),\mathcal{F}_{5}(1,3,5,4,2),\mathcal{F}_{5}(1,4,3,5,2),\mathcal{F}_{5}(1,4,5,3,2),\mathcal{F}_{5}(1,5,3,4,2),\mathcal{F}_{5}(1,5,4,3,2)\right\} .
\end{equation*}

(A) For the factorization on the $s_{12}=q^2$ pole, one has
\begin{equation}
	 ({\vec v}^{A,\sigma}_{5} \cdot \vec{\mathcal{F}}_{5})\big|_{s_{12}=q^2} =\mathcal{F}_{2}(1^{\phi},2^{\phi})\ \mathcal{A}_{5}(\QQ_{2}^{S},3^{g},\sigma\{4^{g},5^{g}\},-q^{S})\,,
\end{equation}
where $\sigma \in \{ \mathbf{1}, \sigma_2 \} = S_2$ and the  two corresponding vectors are 
\begin{align}
{\vec v}^{A,\mathbf{1}}_{5} & = 
\bigg\{ 
\frac{\tau_{31}\tau_{52}}{\tau_{5 q}}\Big(\frac{\tau_{4,(1+3)}}{\tau_{3 \QQ_2}}+1\Big), \frac{\tau_{31}\tau_{42}\tau_{5,(1+3)}}{\tau_{3 \QQ_2}\tau_{5 q}} , -\frac{\tau_{32}\tau_{41}\tau_{52}}{\tau_{3 \QQ_2}\tau_{5 q}}, -\frac{\tau_{32}\tau_{41}\tau_{5,(2+3)}}{\tau_{3 \QQ_2}\tau_{5 q}} , \frac{\tau_{31}\tau_{42}\tau_{51}}{\tau_{3 \QQ_2}\tau_{5 q}} , -\frac{\tau_{32}\tau_{51}}{\tau_{5 q}}\Big(\frac{\tau_{4,(2+3)}}{\tau_{3 \QQ_2}}+1\Big) 
\bigg\} ,  \\
{\vec v}^{A,\sigma_2}_{5} & = 
\bigg\{ 
\frac{\tau_{31}\tau_{52}\tau_{4,(1+3)}}{\tau_{3 \QQ_2}\tau_{4 q}} , \frac{\tau_{31}\tau_{42}}{\tau_{4 q}}\Big(\frac{\tau_{5,(1+3)}}{\tau_{3 \QQ_2}}+1\Big) , \frac{\tau_{31}\tau_{41}\tau_{52}}{\tau_{3 \QQ_2}\tau_{4 q}} ,  -\frac{\tau_{31}\tau_{42}}{\tau_{4 q}}\Big(\frac{\tau_{5,(2+3)}}{\tau_{3 \QQ_2}}+1\Big) , -\frac{\tau_{32}\tau_{42}\tau_{51}}{\tau_{3 \QQ_2}\tau_{4 q}} ,  -\frac{\tau_{32}\tau_{51}\tau_{4,(2+3)}}{\tau_{3 \QQ_2}\tau_{4 q}}
\bigg\} .
\end{align}
One can combine the two vectors to form a 2$\times$6 matrix:
\begin{equation}
\label{eq:bfVmatrix5pt}
{\bf V}^{A}_5=\begin{pmatrix}
{\vec v}^{A,\mathbf{1}}_{5} \\
{\vec v}^{A,\sigma_2}_{5}
\end{pmatrix} ,
\end{equation}
and it appears in the decomposition of the KLT kernel as
\begin{equation}
	 \mathrm{Res}\left[{\bf S}^{\mathcal{F}}_{5}\right]_{s_{12}=q^2}=({\bf V}^{A}_5)^{\rm \scriptscriptstyle T} \cdot \left({\bf S}^{\mathcal{F}}_{2}\otimes{\bf S}^{\mathcal{A}}_{5}\right) \cdot{\bf V}^{A}_5\big|_{s_{12}=q^2}\,.
\end{equation}

(B) For the factorization on the $s_{123}=q^2$ pole, one has 
\begin{equation}
	 ({\vec v}^{B}_{5} \cdot \vec{\mathcal{F}}_{5})\big|_{s_{123}=q^2} =\mathcal{F}_{3}(1^{\phi},3^{g},2^{\phi})\ \mathcal{A}_{4}(\QQ_{3}^{S},4^{g},5^{g},-q^{S})\,,
\end{equation}
where the generalized BCJ vectors are
\begin{equation}
{\vec v}^{B}_5=
\bigg\{
\frac{\tau_{4,(1+3)}\tau_{52}}{\tau_{4 \QQ_3}}, \frac{\tau_{42}\tau_{5,(1+3)}}{\tau_{4 \QQ_3}} , \frac{\tau_{41}\tau_{52}}{\tau_{4 \QQ_3}} , \frac{\tau_{41}\tau_{5,(2+3)}}{\tau_{4 \QQ_3}} , \frac{\tau_{42}\tau_{51}}{\tau_{4 \QQ_3}} , \frac{\tau_{4,(2+3)}\tau_{51}}{\tau_{4 \QQ_3}}
\bigg\} ,
\end{equation}
and it satisfies
\begin{equation}
	 \mathrm{Res}\left[{\bf S}^{\mathcal{F}}_{5}\right]_{s_{123}=q^2}=({\vec v}^{B}_5)^{\rm \scriptscriptstyle T} \cdot \left({\bf S}^{\mathcal{F}}_{3} \otimes {\bf S}^{\mathcal{A}}_{4}\right) \cdot {\vec v}^{B}_5\big|_{s_{123}=q^2}\,.
\end{equation}
Similar results for the $s_{124}=q^2$ or  $s_{125}=q^2$ pole can be obtained by a trivial relabeling.

(C) For the factorization on the $s_{1234}=q^2$ pole, one has
\begin{equation}
	 ({\vec v}^{C,\sigma}_{5} \cdot \vec{\mathcal{F}}_{5})\big|_{s_{1234}=q^2} =\mathcal{F}_{4}(1^{\phi},\sigma\{3^{g},4^{g}\},2^{\phi}) \ \mathcal{A}_{3}(\QQ_{4}^{S},3^{g},-q^{S})\,,
\end{equation}
where $\sigma \in \{ \mathbf{1}, \sigma_2 \} = S_2$ and the  two corresponding vectors are 
\begin{align}
{\vec v}^{C,\mathbf{1}}_{5} & = 
\{ \tau_{52}, \tau_{5,(2+4)}, 0, 0, \tau_{5,(2+3+4)}, 0 \} ,  \\
{\vec v}^{C,\sigma_2}_{5} & = 
\{ 0, 0, \tau_{52}, \tau_{5,(2+3)}, 0, \tau_{5,(2+3+4)} \} .
\end{align}
One can combine the two vectors to form a 2$\times$6 matrix:
\begin{equation}
\label{eq:bfVmatrix5pt-2}
{\bf V}^{C}_5=\begin{pmatrix}
{\vec v}^{C,\mathbf{1}}_{5} \\
{\vec v}^{C,\sigma_2}_{5}
\end{pmatrix} ,
\end{equation}
and it appears in the factorization of the KLT kernel as
\begin{equation}
\mathrm{Res}\left[{\bf S}^{\mathcal{F}}_{5}\right]_{s_{1234}=q^2}=({\bf V}^{C}_5)^{\rm \scriptscriptstyle T} \cdot \left({\bf S}^{\mathcal{F}}_{4}\otimes{\bf S}^{\mathcal{A}}_{3}\right) \cdot {\bf V}^{C}_5\big|_{s_{1234}=q^2}\,.
\end{equation}

We will not give the explicit form of $\Theta_5^{\cal F}$ (and ${\bf S}_{5}^{\cal F}$), but only give its determinant which has the following nice structure as promised in the main text:
\begin{equation}
\det (\Theta_5^{\cal F}) = { \big[ (q^2 - s_{12}) \prod_{3\leq i < j \leq5} (q^2 - s_{12ij}) \big]^2 \prod_{i=3}^5 (q^2 - s_{12i}) \over \big[ s_{1345} s_{2345}\prod_{(ij)\neq(12)} s_{ij} \big]^2 s_{345} \prod_{3\leq i < j \leq5}(s_{1ij} s_{2ij}) } .
\end{equation}
We also point out that the power of the propagators have clear physical meanings and will be explained in \cite{treepaper}. 

\subsection{III. A proof for the general factorization relation \eqref{eq:nptbcjN} for MHV form factors}

\noindent
We consider the generalized BCJ relation \eqref{eq:nptbcjN} in the main text, which are reproduced here:
\begin{align}\label{eq:nptbcjN-AA}
	\mathcal{F}_{n-1}(1,3,\ldots,n-1,2)\ &\mathcal{A}_3(\QQ_{n-1},n,-q)\\
	&= \bigg[ \tau_{n 2}\mathcal{F}_{n}(1,3,\ldots,n,2)+\sum_{i=3}^{n-1}\tau_{n,(2+i + \cdots + (n-1))}\mathcal{F}_{n}{\small (1,3\ldots,i-1, n,i\ldots,n-1,2)} \bigg] \bigg|_{\QQ_{n-1}^2=q^2}\nonumber ,
\end{align}
where $\QQ_{n-1}=\sum_{i=1}^{n-1}p_i$. We will give a recursive proof of this relation for the four-dimensional MHV form factors of $\operatorname{tr}(\phi^2)$, expressed as
\begin{equation}
	{\cal F}_{n}(1^{\phi},\sigma\{3^{+},\ldots,n^{+}\},2^{\phi})=\frac{\langle 12\rangle^2}{\langle 1 \sigma(3)\rangle \cdots \langle \sigma(n) 2\rangle \langle 21 \rangle}\,.
\end{equation}

To prove \eqref{eq:nptbcjN}, we perform a standard BCFW shift, that is the $\langle 2 1]$-shift: $|\hat{2}\rangle=|2\rangle-z |1\rangle$, $|\hat{1}]=|1]+z |2]$. 
We first focus on the LHS, in which only $\mathcal{F}_{n-1}(1,3,\ldots,n-1,2)$ is affected by the shift. We define   
\begin{equation}
	E_{L}(z)=\frac{1}{z}{{\cal F}}_{n-1}({\hat 1},3,\ldots,n-1,{\hat 2};z)\mathcal{A}_{3}(\QQ_{n-1},n,-q)\,,
\end{equation}
so that ${\rm Res}[E_{L}(z)]_{z=0}$ gives the LHS of \eqref{eq:nptbcjN}. Apart from $z=0$, $E_{L}$ have only one other pole on the complex plane $z_{P}=\langle 2 (n-1) \rangle/\langle 1 (n-1) \rangle$, on which the MHV form factors factorize as (here $P=p_2+p_{n-1}$)
\begin{equation}\label{eq:FLzp}
	{\rm Res}[E_{L}(z)]_{z=z_{P}}=-{\cal F}_{n-2}({\hat 1},3,\ldots,\hat{P};z_P)\frac{1}{s_{2(n-1)}}{\cal A}_3({\hat 2},n-1,-\hat{P};z_P) \mathcal{A}_{3}(\QQ_{n-1},n,-q)\,.
\end{equation}
For the RHS of \eqref{eq:nptbcjN}, we can define $E_{R}$ similarly as  
\begin{align}\label{eq:nptbcjN-BB}
	E_{R}(z)=\frac{1}{z}\Big[\tau_{n {\hat 2}}\mathcal{F}_{n}({\hat 1},3,\ldots,n, {\hat 2};z)+\sum_{i=3}^{n-1}\tau_{n,({\hat 2}+i+\cdots+(n-1))}\mathcal{F}_{n}{\small ({\hat 1},3,\ldots,i-1,n,i\ldots,n-1, {\hat 2};z)}\Big] \Big|_{\QQ_{n-1}^2=q^2}\,.
\end{align}
First we point out that the condition $\QQ_{n-1}^2=q^2$ will not be spoiled by the shift since $\hat{p}_1+\hat{p}_2=p_1+p_2$. Next we examine the possible poles of $E_{R}(z)$. For the form factors in the sum, only $z_{P}$ pole appears. For the first form factor $\mathcal{F}_{n}({\hat 1},3,\ldots,n, {\hat 2};z)$, naively one may expect a pole $\langle 2 n\rangle /\langle 1 n\rangle$; however, it is canceled by the $\tau_{n\hat{2}} = \langle 2 n \rangle [n 2] - z \langle 1 n \rangle [n 2]  $ factor. Moreover, we need to be careful about the pole at infinity: although the MHV form factors themselves do not contribute to the pole at infinity, the $\tau$ factors do. 
One can compute the corresponding residue  for each term in $E_{R}(z)$ as 
\begin{equation}
	{\rm Res}\big[\tau_{n,(\hat{2}+\cdots)}\mathcal{F}_n({\hat 1},\sigma\{3,\ldots,n\},{\hat 2})\big]_{z=\infty}=\frac{\langle 1 n \rangle [ n 2 ] \langle 2 1 \rangle}{\langle 1 \sigma(3) \rangle \cdots \langle\sigma(n) 1 \rangle }\,. 
\end{equation}
where $\sigma$ can be arbitrary permutations, and it turns out that the sum of the residues actually vanishes: 
\begin{equation}
{\rm Res}[E_{R}(z)]_{z=\infty}= \sum_{i=3}^{n} \frac{\langle 1 n \rangle [ n 2 ] \langle 2 1 \rangle}{\langle 1 3 \rangle \cdots \langle (i-1)  n \rangle \langle n i \rangle \cdots \langle (n-1) 1 \rangle }=0\,,
\end{equation}
which is equivalent to an $(n-1)$-point U(1) decoupling relation.
Therefore, we find that  $E_{R}$ have also only the $z_{P}$ pole (apart from the $z=0$ one). And the residue is 
\begin{equation}\label{eq:FRzp}
	{\rm Res}[E_{R}(z)]_{z=z_{P}}=-\sum_{i=3}^{n-1}\tau_{n,(\hat{P}+i+\cdots+(n-2))}{\mathcal{F}}_{n-1}{\small ({\hat1},3, \ldots,i-1,n,i\ldots,{\hat P};z_P)}\frac{1}{s_{2(n-1)}}{\cal A}_3({\hat 2},n-1,-\hat{P};z_P)\big|_{\QQ_{n-1}^2=q^2}\,.
\end{equation}

Comparing \eqref{eq:FLzp} and \eqref{eq:FRzp}, one can see that ${\rm Res}[E_{L}(z)]_{z=z_{P}}={\rm Res}[E_{R}(z)]_{z=z_{P}}$ 
by using a $(n-1)$-point relation \eqref{eq:nptbcjN-AA}, and the residue theorem guarantees that ${\rm Res}[E_{L}(z)]_{z=0}={\rm Res}[E_{R}(z)]_{z=0}$, so that  \eqref{eq:nptbcjN-AA} is valid for the $n$-point case. 

\subsection{IV. Further detail about $n$-point double copy and hidden factorization relations}
\noindent
Here we discuss detailed notations of the generic $n$-point form factor double copy and hidden factorization relations.

We start from the generalized factorization relations \eqref{eq:generalfactorization} for the generic $n$-point case which reads in a more precise form as
\begin{equation}\label{eq:nptgeneralizedBCJ}
		\sum_{\alpha\in S_{n-2}} {\vec v}_{(\bar{\kappa},\bar{\rho})}[\alpha]\mathcal{F}_{n}[\alpha]\big|_{\QQ_m^2=q^2}= \mathcal{F} _{m}[\bar{\kappa}]\mathcal{A} _{m^{\prime}}[\bar{\rho}] \equiv \left(\mathcal{F} \mathcal{A}\right)_{(m,m^{\prime})}[\bar{\kappa},\bar{\rho}]\,,
\end{equation}
with $\QQ_{m}=\sum_{i=1}^{m}p_i$ and  $m^{\prime}=n-m+2$. Here $ \mathcal{F} _{m}[\bar{\kappa}]$ and $\mathcal{A} _{m^{\prime}}[\bar{\rho}]$ are the color-ordered $m$-point form factors and $m'$-point amplitudes defined explicitly as
\begin{equation}
\begin{aligned}
	&\mathcal{F} _{m}[\bar{\kappa}]\equiv \mathcal{F} _{m}(1, \bar{\kappa}\{3,\ldots,m\},2),\quad \bar{\kappa} \in S_{m-2};\\
	& \mathcal{A} _{m^{\prime}}[\bar{\rho}]\equiv \mathcal{A} _{m^{\prime}}(\QQ_m,m+1, \bar{\rho}\{m+2,\ldots,n\},-q), \quad \bar{\rho} \in S_{m'-3}\,.
\end{aligned}
\end{equation}
For the form factors, we use the DDM basis, and for amplitudes, we use the BCJ basis with $\QQ_{m}, q$ and $(m+1)$ (adjacent to $\QQ_m$) fixed.  From \eqref{eq:nptgeneralizedBCJ}, we see that the generalized BCJ vectors ${\vec v}_{(\bar{\kappa},\bar{\rho})}$ depend on the ordering $(\bar{\kappa},\bar{\rho})$ of the basis amplitudes and form factors, and each of them is assigned a generalized BCJ relation.  

To relate \eqref{eq:generalfactorization} to the gravitational factorization property as in \eqref{eq:nptfactorizationGRA}, one  observe that the same vectors ${\vec v}_{(\bar\rho,\bar\kappa)}[\alpha]$ also induce a decomposition of the KLT matrix ${\bf S}^{\cal F}_{n}$:
\begin{align}\label{eq:nptSFdecomposition}
	&{\rm Res}\left[\mathbf{S}^{\mathcal{F}}_{n}[\alpha_1|\alpha_2]\right]_{\QQ_{m}^2=q^2}=\sum_{\substack{\bar{\kappa}_{1,2}\in S_{m-2} \\ \bar{\rho}_{1,2}\in S_{m^{\prime}-3} } } {\vec v}_{(\bar\kappa_1,\bar\rho_1)}[\alpha_1]\left(\mathbf{S}^{\mathcal{F}}_{m}[\bar{\kappa}_1|\bar{\kappa}_2] \otimes \mathbf{S}^{\mathcal{A}}_{m^{\prime}}[\bar{\rho}_1|\bar{\rho}_2]\right) {\vec v}_{(\bar\kappa_2,\bar\rho_2)}[\alpha_2]\,. 
\end{align}
This equation can be regarded as a matrix product, as shown in the schematic equation \eqref{eq:NptSfdecomp} in the main text, and the collection of ${\vec v}_{(\bar\rho,\bar\kappa)}[\alpha]$ form a $(m-2)!(m'-3)!\times (n-2)!$ matrix denoted as ${\bf V}$ in \eqref{eq:NptSfdecomp}. Explicit examples can be found in the previous five-point examples such as \eqref{eq:bfVmatrix5pt} and \eqref{eq:bfVmatrix5pt-2}.

A nice consequence of  \eqref{eq:nptgeneralizedBCJ} and \eqref{eq:nptSFdecomposition}  is the factorization formula of the gravity theory \eqref{eq:nptfactorizationGRA}:
\begin{equation}\label{eq:tensorproductfactorize}
\begin{aligned}
{\rm Res}\left[\mathcal{G}_{n}\right]_{\QQ_{m}^2=q^2}&=
\sum_{\alpha_{1,2}} \mathcal{F} _{n}[\alpha_1] {\rm Res}\left[ \mathbf{S}^{\mathcal{F}}_{n}\right][\alpha_1|\alpha_2] \mathcal{F} _{n}[\alpha_2]\big|_{\QQ_{m}^2=q^2}\\
	&=\sum_{\bar{\kappa},\bar{\rho}}
	 \left(\mathcal{F} \mathcal{A}\right)_{(m,m^{\prime})}[\bar{\kappa}_1,\bar{\rho}_1] 
	\left(\mathbf{S}^{\mathcal{F}}_{m}[\bar{\kappa}_1|\bar{\kappa}_2] \otimes \mathbf{S}^{\mathcal{A}}_{m^{\prime}}[\bar{\rho}_1|\bar{\rho}_2]\right)
	 \left(\mathcal{F} \mathcal{A}\right)_{(m,m^{\prime})}[\bar{\kappa}_2,\bar{\rho}_2]\\
	 &=
	 \sum_{\bar{\kappa}_{1,2}} 
	 \mathcal{F}_{m}[\bar{\kappa}_1]\mathbf{S}^{\mathcal{F}}_{m}[\bar{\kappa}_1|\bar{\kappa}_2]\mathcal{F}_{m}[\bar{\kappa}_2]
	 \sum_{\bar{\rho}_{1,2}} 
	 \mathcal{A}_{m^{\prime}}[\bar{\rho}_1]\mathbf{S}^{\mathcal{A}}_{m^{\prime}}[\bar{\rho}_1|\bar{\rho}_2]  \mathcal{A}_{m^{\prime}}[\bar{\rho}_2]= \mathcal{G}_{m}\mathcal{M}_{m'} \,,
\end{aligned}
\end{equation}
which is a detailed $n$-point generalization of \eqref{eq:4ptFactorization} in the paper.

We make a remark that the derivation \eqref{eq:tensorproductfactorize} shows that once we acknowledge the validity of two properties: (1) the generalized BCJ relation \eqref{eq:nptgeneralizedBCJ}, which is practically easier to check in gauge theories, and (2) the decomposition of the KLT kernel \eqref{eq:nptSFdecomposition}, which is irrelevant to the specific theory or operator, there is no need to worry about the factorization on the new $\QQ_{m}^2=q^2$ poles exposed by double copy. This is a crucial step to confirm that ${\cal G}_{n}$ is indeed a physically meaningful quantity in gravity. 

Other physical requirements are easier to address, such as the manifest diffeomorphism invariance and factorization properties on the physical poles, \emph{i.e.}~those poles appearing already in gauge form factors. To argue the latter point, it is more convenient to look at an alternative form of the KLT double copy 
\begin{equation}\label{eq:genericKLT2}
\mathcal{G}_{n} =\sum_{\alpha,\beta\in S_{n-2}}{N} _{n}[\alpha] \, {\Theta}^{\mathcal{F}}_{n}[\alpha|\beta] \, {N} _{n}[\beta]\,. 
\end{equation}
Since ${\Theta}^{\mathcal{F}}_{n}$ can be understood as the form factor in the bi-adjoint scalar theory, it also has a factorization on, say the $P^2_{m}=s_{(1 i_1\ldots i_m)}=0$ pole, 
so that $\mathrm{Res}\left[{\Theta}^{\mathcal{F}}_{n}\right]_{P_m^2=0} = {\bf U}^{\rm \scriptscriptstyle T} \cdot ({\Theta}^{\mathcal{A}}_{m+2} \otimes {\Theta}^{\mathcal{F}}_{n-m}) \cdot {\bf U}$.  
Here ${\bf U}$ is the orthogonal complement of BCJ vectors for amplitudes, relevant only to the $\Theta^{\cal F}$ block.
The numerators can also factorize schematically ${\bf U} \cdot {\vec N}_{n}\big|_{P_{m}^2=0} = N_{m+2} N_{n-m}$.
Thus, starting from \eqref{eq:genericKLT2} and performing some refinements similar to \eqref{eq:tensorproductfactorize}, we have 
\begin{equation}\label{eq:generalphysical}
	\mathrm{Res}\left[{\mathcal{G}_{n} }\right]_{P_m^2=0}=\big(N_{m+2}{\Theta}^{\mathcal{A}}_{m+2}N_{m+2}\big)\times\big(N_{n-m}{\Theta}^{\mathcal{F}}_{n-m}N_{n-m}\big)=\tilde{\mathcal{M}}_{m+2} \mathcal{G}_{n-m} \,,
\end{equation}
here $\tilde{\mathcal{M}}$ refers to gravity amplitudes with only massless scalar $\phi$ and gravitons. We finally comment that a similar analysis with exactly the same ${\bf U}$ applies to amplitudes, which conversely increases our confidence on \eqref{eq:generalphysical} for form factors.

\end{document}